\documentclass[pra,onecolumn,eqsecnum,superscriptaddress,preprint]{revtex4}
\usepackage{epsfig}

\begin{document}

\title{Statistical mechanics of Floquet systems: the pervasive problem of near degeneracies}

\author{Daniel W.~Hone}
\email{hone@kitp.ucsb.edu}
\affiliation{Physics Department, UCSB, Santa Barbara, CA 93106}
\affiliation{Kavli Institute for Theoretical Physics, UCSB, 
Santa Barbara, CA 93106}

\author{Roland Ketzmerick}
\affiliation{Institut f\"ur Theoretische Physik, 
Technische Universit\"at Dresden, 01062 Dresden, Germany}

\author{Walter Kohn}
\email{kohn@physics.ucsb.edu}
\affiliation{Physics Department, UCSB, Santa Barbara, CA 93106}

\date{\today}

\begin{abstract}
Although the statistical mechanics of periodically driven (``Floquet") systems
in contact with a heat bath
has some formal analogy with the traditional statistical mechanics of
undriven systems, closer examination reveals radical differences. 
In Floquet systems {\it all} quasienergies $\varepsilon_j$ can be placed in a finite
frequency interval $0\le \varepsilon_j <\omega$ (with $\omega$ the driving
frequency and $\hbar =1$), and the number of near degeneracies 
($|\varepsilon_j-\varepsilon_k|\le\delta$ for arbitrarily small $\delta$) 
in this interval grows
without limit as the dimension $N$ of the Hilbert space increases. 
As we noted in a previous paper, this leads to pathologies, including drastic changes
in the Floquet states, as $N$ increases.  
In earlier work on Floquet systems in contact with
a heat bath these difficulties were put aside by fixing 
$N$, while taking the coupling to the bath to be smaller than any quasienergy
difference.  This led to a simple
explicit theory for the reduced density matrix, but with some 
major differences from the usual time independent statistical mechanics.
We show that, for weak but finite
coupling between system and heat bath, the accuracy of a calculation within the truncated
Hilbert space spanned by the $N$ lowest energy eigenstates of the undriven
system is limited, as $N$ increases indefinitely, only by the usual neglect of
bath memory effects within the Born and Markov approximations.
As we seek higher accuracy by increasing $N$,
we inevitably encounter quasienergy differences {\it smaller} than the system-bath coupling. 
We therefore derive the steady state reduced density matrix without restriction on
the size of quasienergy splittings. In general, it is no longer diagonal 
in the Floquet states.  We analyze, in particular, the
behavior near a weakly avoided crossing, where quasienergy near degeneracies routinely appear.
The explicit form of our results for the denisty matrix gives a consistent prescription for the
statistical mechanics for many periodically driven systems with $N$ infinite, in spite of the 
Floquet state pathologies.
\end{abstract}

\maketitle

\newpage
\section{Introduction}
Time-periodic quantum systems are of great importance in several fields
of science:   atomic and solid-state systems driven by
monochromatic electromagnetic fields~\cite{holthaus}; quantum ratchets~\cite{ratchets};
quantum chaos~\cite{quantumchaos}.
The Floquet theorem gives the solutions of the
time-dependent Schr\"odinger equation with a time-periodic 
Hamiltonian (period $2\pi/\omega$) in the form 
$\exp(-i\varepsilon t)u(t)$, with $u(t)$ exhibiting
the same periodicity as the Hamiltonian.
The ``quasienergy" $\varepsilon$ plays a role analogous to the energy
in static systems, determining in this case the behavior of the wave function
under finite one-period, rather than infinitesimal, time translations.
The quasienergy $\varepsilon$ is defined only mod $\omega$, and we
follow the convention of taking $0\leq\varepsilon <\omega$. 

Real Floquet systems will ordinarily be in thermal contact with a
bath.  Energy is being continuously supplied by the periodic driving field,
with relaxation provided by the bath.  This leads to a fundamental
question of statistical physics:  Is there an asymptotic, time
periodic,  ensemble characterizing the quantum system and,
if so, how does it 
depend on the coupling to and the temperature of the bath?
For a time {\it independent} system with Hamiltonian $H_0$ the answer is
given by the canonical reduced density operator $\rho = \exp(-\beta H_0)/Z$,
where $\beta$ is the inverse temperature, and
$Z$ is the partition function.  This density operator is
independent of the form and magnitude of the coupling to the environment, 
to lowest nontrivial order in that coupling,
and the operator is diagonal in the energy eigenstates of
the system Hamiltonian.  Thus, even if two states are separated by
an energy small compared to their coupling with the bath 
(``near degeneracy"), or even if they are degenerate, their statistical weights are
given by their energy, and there is no anomaly associated with their near or
strict degeneracy. That simplicity does {\it not} carry over to the case of
near degeneracy in {\it quasienergy} .

For time-periodic systems in thermal contact with a bath the general long time
behavior has not yet been fully clarified.  In particular, the inevitable appearance
of near degeneracies of quasienergies has not been addressed.  Earlier work includes
a number of papers~\cite{blumel,doh,graham,kohler,grhan,breuer} which
have discussed specific physical systems, treating either those whose quantum states
span only a finite space, or others whose Hilbert space has been truncated to a
finite set and then studied numerically, but always with coupling to the bath small
with respect to any quasienergy spacing.  Within such a finite state space,
and in the limit of weak coupling to the bath, a description of  driven
dissipative systems, as applied to driven Rydberg atoms in thermal
contact with discrete wave guide modes, 
was introduced by Bl\"umel {\it et al.}~\cite{blumel}, and extended by
Kohler {\it et al.}~\cite{kohler} to driven one-dimensional harmonic and anharmonic oscillators.
An extension to strong coupling to the bath was presented by
Thorwart {\it et al.}~\cite{thorwart} for tunneling and relaxation of a particle
in a bistable one-dimensional well, but using methods and approximations
suitable only when a (small) finite number of system energy eigenstates
is significantly accessible.
A  discussion of the statistical mechanics of a more general class of periodically
driven systems was given by one of us~\cite{kohn}, restricted 
to a finite Hilbert space with a non-degenerate quasienergy spectrum, in 
the limit of coupling to the bath much weaker than the smallest quasienergy
spacing.  In this limit, where the density
operator is assumed diagonal in the representation of the Floquet states, the matrix
elements can be obtained by solving a set of linear rate equations.
In sharp contrast to the time independent case, they do depend on the
explicit form (but not the magnitude) of the coupling between system and
bath, and they are not simple smooth functions of the quasienergy.

In the general case of time-periodic quantum systems with a
Hilbert space whose number of dimensions $N$ increases
without limit,  the situation is more complex. Since the 
quasienergies are restricted to a finite range of width $\omega$,
the mean spacing between neighboring quasienergy levels approaches
zero as their number increases.  
Therefore, any fixed coupling to the thermal environment is never small relative to
all level spacings for a sufficiently large basis size.  It becomes essential
to deal with degenerate and near degenerate quasienergies, a situation which has been
explicitly excluded in previous analyses.   Indeed, in the limit $N\rightarrow\infty$ 
an infinite number of quasienergy degeneracies
avoided by arbitrarily weak level repulsions may occur, and many 
anomalies of the spectrum and the Floquet states appear, as was 
pointed out in~\cite{hkk}.  So it is of special importance to determine
the effect of the bath when the coupling becomes comparable to or even
large with respect to one or more of the relevant quasienergy splittings.

In this paper we address the critical question of the appropriate description 
of the steady state of time-periodic quantum systems with small but finite coupling to the bath.
We review in Sec. II the derivation from the master equation of the set of linear
equations that determine the reduced density matrix.  This
matrix is in general no longer diagonal in the basis of the  Floquet states of the system,
but it does approach a limiting time periodic form. 
We obtain a set of linear homogeneous equations which
determine all of the density matrix elements 
in the long time, stationary time-periodic state, Eq.~(\ref{rate_equation_general}).
This result is no longer limited to the regime of bath coupling weak compared
to all quasienergy spacings.
%was previously discussed~\cite{blumel,kohler}
%only for situations where all quasienergy spacings were large compared with the coupling
%to the bath (and therefore only for systems restricted to a finite number of states).
For a finite Hilbert space and in the limit of infinitesimal coupling 
we discuss the differences from the conventional
statistical mechanics of time-independent systems.

We next study in detail the influence  of a single weakly avoided crossing of 
Floquet quasienergies (as a function of driving field strength) 
on a finite density matrix, since it is there
that the near degeneracies of interest develop.  In Sec. III we find that the
important consequences of such an avoided crossing can be incorporated by introducing 
an additional effective coupling~(\ref{defRAC}) between the two
states involved, mediated by the bath.  This coupling decreases
with distance from the avoided crossing (which conveniently parameterizes the
quasienergy separation of the two states), and it depends on the ratio of 
the coupling strength $\Gamma$ of the two states to the bath, 
and their minimum quasienergy splitting, $\Delta$.  Remarkably, all of the diagonal
elements of the density matrix in the representation of the ``diabatic" states
defined well away from the center of the avoided crossing,  a basis which is  fixed
throughout the crossing, are determined by a set of linear 
rate equations (\ref{ac3}) of exactly the structure familiar from ordinary 
time-independent statistical mechanics, but with this single added effective
coupling rate.  For $\Gamma\ll\Delta$ the additional coupling  
influences not only
the (diagonal and off-diagonal) reduced density matrix elements
for the two states themselves, but may completely change the whole
steady state reduced density matrix, so that average system observables will in general
change radically with driving field strength in the range of  
the avoided crossing.  Physically, high order near resonances of the driving field
between eigenstates of the undriven system can populate high energy levels, which
then affect populations of lower energy levels in the relaxation cascade. 
In the opposite limit, $\Gamma\gg\Delta$, the coherence necessary for the high
order resonance is disrupted by the interactions with the bath, and the avoided crossing has 
negligible influence on the reduced density matrix which, in this limit, is approximately
diagonal in the diabatic basis.  We present a simple, albeit somewhat artificial,  
example which can be solved in some detail, specifically exhibiting these effects.

We then use this insight to discuss time-periodic quantum systems with a 
Hilbert space of increasingly large dimension in Sec. IV.  
We find that weakly enough avoided crossings, while having great impact
on the Floquet states, have no significant consequences for the density
matrix (in a $\lambda$-independent basis).  
It is this central result which justifies the neglect in many cases of all but a finite basis set
in the calculation of the statistical mechanics of a Floquet system, as is essential
for any numerical study.  Specifically, for the calculation
of statistical properties, we establish the validity of using the conventional
approximation of eliminating from the {\it undriven} system spectrum all states of
sufficiently high energy (the cutoff to be determined by the accuracy desired), and
allowing the periodic driving field to mix only the remaining states.  Within that finite 
basis set we have derived a prescription for calculating the full reduced density matrix, 
in general no longer diagonal in the Floquet representation.  A brief summary is given 
in the conclusion, Sec. V.

\section{Reduced Density Matrix}
\subsection{Equations of motion and time periodic solutions}

We review and extend the derivation of the steady state solution for
the reduced density matrix in the weak-coupling limit (``Born-Markov approximation") 
for Floquet systems, previously discussed~\cite{blumel,kohler}
only for situations where all quasienergy spacings were large compared with the coupling
to the bath  (and therefore only for systems restricted to a finite number of states).
In contrast, we stress that the validity of the more general equations is 
not limited to that regime, and we give a detailed justification for an approximation using
time-averaged rates.

We write the Hamiltonian generally as the sum of $H_s(t)$, describing
the dynamics of the system driven by a time-periodic external
field with period $2\pi/\omega$, the Hamiltonian $H_b$ of the bath, 
and the interaction between system and bath $H_{sb}$,
\begin{equation}
H(t) = H_s(t) + H_b  + H_{sb} .
\label{Hfull}
\end{equation}
The total density operator of this closed  coupled system, $\rho_T(t)$, is the
solution of the equation of motion, 
\begin{equation}
i\dot\rho_T(t) = [H(t),\rho_T(t)] .
\label{rhot}
\end{equation}
%(We have set $\hbar=1$ for simplicity.)
We trace over bath variables to obtain the reduced density operator
for the system,
\begin{equation}
\rho = {\rm Tr_b}\,\rho_T,
\end{equation} 
and do the standard recasting of the master equation into an integro-differential equation 
(see, e.g., \cite{blum,ct}), whose iteration gives directly the expansion in powers
of the system-bath interaction:
\begin{equation}
\frac{\rm d}{\rm dt}\tilde\rho(t) = -\int_0^t dt' {\rm Tr_b}
\left[\tilde H_{sb}(t),\left[\tilde H_{sb}(t'),\tilde\rho_T(t')
\right]\right].
\label{exactint}
\end{equation}
Here the tilde labels operators in the interaction picture.  We have omitted 
the linear term involving  ${\rm Tr_b}H_{sb}$, which ordinarily vanishes.  
This is the case, e.g., for
an interaction between system and bath characterized by creation or
destruction of single bath bosons, such as phonons or photons, or of free
electrons or other fermions.  Otherwise, this equation is still general and exact.

In order to make further progress, which depends on being able to
trace over the bath variables under the integral, we make two 
essential simplifications: (i) second order perturbation theory \cite{born} in $H_{sb}$ 
and (ii) the Markov approximation.
Their validity relies ultimately on being able to neglect any memory
effects within the bath.  The correlation times associated with the
bath must be small compared with the times characterizing the evolution
of the reduced density operator (see the condition
(\ref{bornmarkovcondition}) below).  
First, to lowest non-vanishing order in $H_{sb}$ we take the total
density operator $\rho_T(t')$ under the integral as a simple product of system and bath factors,
\begin{equation}
\tilde\rho_T(t') \approx  \tilde\rho(t')\rho_b .
\label{born}
\end{equation}
The physical meaning of this approximation is that we neglect more than a single
interaction of the system with the bath during the bath correlation time. 
The effect of higher order correlations, where subsequent interactions depend on
the state of the bath modified by earlier ones, 
is negligible, as we show in more detail in Appendix~\ref{4thorder}.

We make the specific, but inessential, choice of coupling
Hamiltonian:
\begin{equation}
\label{AR}
H_{sb} = \gamma A B ,
\end{equation}
where the Hermitian operator $A$ acts only on the Floquet system,
and the Hermitian operator $B$ acts only on the heat bath. A sum of
such factorized operator products, $\gamma\sum_\alpha A_{\alpha}B_{\alpha}$,
is also readily treated.   But if the various bath operators 
$B_{\alpha}$ are dynamically uncorrelated with each other,
then the results below are modified only by an extra index on
each matrix element of $A$ and each spectral function $g(E)$
(to be defined below),
and a sum over that index.  We take the single product to avoid
the consequent algebraic clutter, since no essential difference
results in the ultimate equations (the extra sums being merely
incorporated into the definition of the complex ``rates" below
(\ref{time_complexrates})). 
Then the trace over bath variables involves
only the single combination
\begin{equation}
\label{Gdef}
G(\tau) = {\rm Tr}_b \rho_b\tilde B(t)\tilde B(t-\tau),
\end{equation}
This bath correlation function is characterized by a decay time
$\tau_c$, which we assume to be much shorter than the times
characterizing the evolution of the reduced density operator
$\tilde\rho(t)$ (see Eq.~(\ref{bornmarkovcondition}) below). Then the
main contribution to the integral in (\ref{exactint}) is from
values of $t'$ within a range of order $\tau_c$ below the upper
limit $t$.  Thus to a good approximation we can extend the lower 
limit to $-\infty$ and also make the Markov approximation:
replace $t'$ by $t$ in the argument of $\tilde\rho_T$ under
the integral. Returning to the Schr\"odinger picture and 
changing the integration variable to $\tau = t-t'$, we
then have 
\begin{equation}
\label{our4.27}
\frac{{\rm d}\rho}{{\rm d}t} + i [H_s(t), \rho(t)] =
- \gamma^2 \int_0^{\infty} {\rm d}\tau 
\left\{ G(\tau) \left[ A \tilde A(t-\tau ,t) \rho(t)
			- \tilde A(t-\tau ,t) \rho(t) A \right]
			+ {\rm H.c.} \right\}  .
\end{equation}
The rate of change of the reduced density operator due to the coupling
to the bath is of order
\begin{equation}
\Gamma_2 = \gamma^2 \tau_c \langle A^2\rangle \langle B^2\rangle ,
	\label{defgamma2}
\end{equation}   
suggesting that the condition for validity of the ``Born" and 
Markov approximations is
\begin{equation}
	\tau_c \Gamma_2 \ll 1 .
	\label{bornmarkovcondition}
\end{equation}
Appendix~\ref{4thorder} demonstrates that this same inequality assures that higher
order corrections are negligible. 
The validity of the above approximations is discussed in the literature on
undriven systems; a particularly clear discussion is given in  Ref.~\cite{ct} .
As the approximations have been made within the interaction picture, they are 
equally valid for a time-periodic system Hamiltonian $H_s(t)$.  We emphasize that
we do {\it not} require infinitesimal coupling to the bath.  That coupling
need only be small enough to satisfy the condition (\ref{bornmarkovcondition}).

We recall that the solutions of the isolated system Schr\"odinger equation 
$i \partial_t|\phi\rangle = H_s(t) |\phi\rangle$
are of the Floquet form
$|\phi_j(t)\rangle = \exp(-i\varepsilon_j t) |u_j(t)\rangle $
with the quasienergies $\varepsilon_j$ 
chosen to lie in the strip $0\leq\varepsilon <\omega$
and with the
time-periodic parts $|u_j(t)\rangle = |u_j(t+ 2\pi/\omega)\rangle$ 
forming a complete orthonormal basis for the system {\it at any given time $t$}. 
We introduce the matrix elements of the 
reduced density operator in this time-periodic basis,
\begin{equation}
\label{rhoijt}
\rho_{ij}(t) = \langle u_i(t) | \rho(t) | u_j(t) \rangle  ,
\end{equation}
which will turn out to be very convenient in the following calculations
and which, as we will show later, leads to time-independent density matrix
elements in steady state.
In this basis the equation of motion of the reduced density matrix elements takes
the form
\begin{eqnarray}
\label{gterms}
(\frac{\partial}{\partial t} +i \varepsilon_{ij}) \rho_{ij}(t) =
- \gamma^2 \sum_{k,l,M,m} e^{i(M+m)\omega t} \Bigl[ 
&+& A_{ik}(M) A_{kl}(m) \rho_{lj}(t) \int_0^{\infty} {\rm d}\tau G(\tau)
e^{i(\varepsilon_{lk}-m\omega)\tau} \\
&-& A_{ik}(m) \rho_{kl}(t) A_{lj}(M) \int_0^{\infty} {\rm d}\tau G(\tau)
e^{i(\varepsilon_{ki}-m\omega)\tau} \nonumber \\
&+& \rho_{ik}(t) A_{kl}(m) A_{lj}(M) \int_0^{\infty} {\rm d}\tau G^*(\tau)
e^{i(\varepsilon_{lk}-m\omega)\tau} \nonumber \\
&-& A_{ik}(M) \rho_{kl}(t) A_{lj}(m) \int_0^{\infty} {\rm d}\tau G^*(\tau)
e^{i(\varepsilon_{jl}-m\omega)\tau} \nonumber 
\Bigr] ,
\end{eqnarray}
where we defined the Fourier components of the matrix elements of the system operator $A$, 
\begin{equation}
\label{Aij}
\langle u_i(t) | A | u_j(t) \rangle  \equiv  \sum_m A_{ij}(m) e^{im\omega t},
\end{equation}
with the property $A_{kl}(m) = A_{lk}^*(-m)$,
and where we have introduced the notation
$\varepsilon_{ij} \equiv \varepsilon_i -\varepsilon_j$.

We simplify the notation by introducing complex time-periodic ``rates"
\begin{equation}
\label{time_complexrates}
R_{ij;kl}(t) = \sum_K e^{iK\omega t} R_{ij;kl}(K),
\end{equation}
with Fourier coefficients
\begin{equation}
\label{Km_complexrates}
R_{ij;kl}(K) = \sum_m R_{ij;kl}^m(K) = 2\pi\gamma^2 
             \sum_m A_{ij}(m+K) A_{kl}^*(m) g(\varepsilon_{lk}-m\omega)
,
\end{equation}
where $g(E)$ is the Fourier transform of $G(\tau)$. We will approximate
\begin{equation}
\int_0^{\infty} {\rm d}\tau G(\tau)e^{-iE\tau} \approx \pi g(E) ,
\label{prinval}
\end{equation}
neglecting a principal value contribution. From the definition 
of $G(\tau)$, Eq.~(\ref{Gdef}), one finds
\begin{equation}
g(E) = \sum_{\nu,\nu'} P_{\nu} |\langle \nu | B | \nu' \rangle |^2
\delta (E+E_{\nu'}-E_{\nu}) ,
\label{gE}
\end{equation}
where $\nu$ labels bath states of energy $E_{\nu}$
which are in thermal equilibrium and have occupation 
$P_{\nu} \propto \exp(-\beta E_{\nu})$.

Then we have as the fundamental equation of motion
of the reduced density matrix elements in the Floquet representation,
the following set of linear differential equations:
\begin{equation}
\label{time_rate_equation_general}
(\frac{\partial}{\partial t} +i \varepsilon_{ij}) \rho_{ij}(t) =
- \frac{1}{2} \sum_{k,l} \Bigl[
    \rho_{lj}(t) R_{ik;lk}(t) + \rho_{il}(t) R_{jk;lk}^*(t)
   -\rho_{kl}(t) (R_{lj;ki}(t) + R_{ki;lj}^*(t)) \Bigr] .
\end{equation}

This is the generalized Master equation for all density matrix elements.
We next replace the time periodic rates $R_{lj;ki}(t)$
by their average over one driving period,
\begin{equation}
\label{averagerates}
R_{lj;ki} =  \sum_m R^m_{lj;ki} =  R_{lj;ki}(K=0).
\end{equation}
This approximation~\cite{kohler} implicitly assumes that the density matrix elements
do not vary substantially over a period of the driving, and it
is further discussed in Appendix~\ref{validitytimeaverage}. 
We then find a steady state solution of (\ref{time_rate_equation_general}) 
for which the matrix elements $\rho_{ij}$ are independent of time: 
\begin{equation}
\label{rate_equation_general}
i \varepsilon_{ij} \rho_{ij} =  - \frac{1}{2} \sum_{k,l} \Bigl[
\rho_{lj} R_{ik;lk} + \rho_{il} R_{jk;lk}^* 
- \rho_{kl} ( R_{lj;ki} + R_{ki;lj}^*) \Bigr] .
\end{equation}

\subsection{Large quasienergy splittings}

The states of some systems of interest span only a finite dimensional space
(e.g., spin systems).
Because of the finite number of states,
at a given driving strength $\lambda$ there will be a smallest
quasienergy separation. 
In this subsection we take that smallest separation to be large compared to the 
coupling to the heat bath (an assumption made, in fact, in all earlier work
on driven dissipative systems in the weak-coupling limit), and 
we will point to both the similarities to and the differences from the conventional 
statistical mechanics of time-independent systems.  

In the limit where the absolute values of all ``rates" $|R_{ij;kl}|$
are sufficiently small compared with {\it all} splittings 
$|\varepsilon_{ij}|$, it
is clear from (\ref{rate_equation_general}) that all 
off-diagonal density matrix 
elements can be neglected, and we arrive at the familiar form,
\begin{equation}
\label{diagonal}
0 = \rho_{ii} \sum_k R_{ik} - \sum_k \rho_{kk} R_{ki},
\end{equation}
relating the state occupation probabilities $p_i\equiv\rho_{ii}$.
Here the diagonal time-independent rates,
\begin{equation}
\label{realrates}
R_{ki} = R_{ki;ki} = 2\pi\gamma^2 \sum_m |A_{ki}(m)|^2 
g(\varepsilon_{ik} - m\omega) ,
\end{equation}
are of the form of the usual transition rates
of perturbation theory, the square of coupling matrix elements
times the density of states (at the energy transfered from bath
to system), summed over initial states of the
bath with appropriate thermal equilibrium weighting.
Of course, in the static case the equation also holds without the
summation over $k$ (detailed balance).  That is {\it not} the case here.

As was pointed out earlier~\cite{kohn,breuer}, there
are some important differences in this result from that familiar
in static problems.  Because of the possibility of absorption and
emission of multiple quanta of the time-dependent driving field, the
occupation probabilities are not simple Boltzmann factors, and indeed
they depend on the detailed form of the Hamiltonian coupling the 
system to the heat bath.  From the definition (\ref{gE}) of the
spectrum of the bath correlation function we have
\begin{equation}
g(E) = g(-E) e^{-\beta E} ,
\end{equation}
which is used in the time-independent case to derive the detailed balance
equation. In a Floquet system it leads,
using (\ref{gE}) and (\ref{realrates}), to the relation
\begin{equation}
R_{jj'}^m = R_{j'j}^{-m} e^{\beta (\varepsilon_j - \varepsilon_{j'} + m \omega)} 
,
\end{equation}
relating only the pieces of the transition rates between system states
$j$ and $j'$ which involve the transfer of exactly $m$ quanta from
the field.  After summation over $m$ we find for the ratio of the
full transition rates between these states,
\begin{equation}
\label{ratioofrates}
\frac{R_{jj'}}{R_{j'j}} = \frac{\sum_m R_{jj'}^m}{e^{-\beta (\varepsilon_j
-\varepsilon_{j'})} \sum_m R_{jj'}^m e^{-\beta m \omega}} .
\end{equation}

There are situations where the $R_{jj'}^m$ 
have for some pairs $(j,j')$ just one value of $m$,
say $m_0$, with a non-negligible value.
This is the case for
a very small driving amplitude ($\lambda \ll 1$)
or for high-lying states of the undriven system that are weakly
affected by the periodic driving,
if the levels $j$ and $j'$ are far enough from any avoided 
crossing. Then in the above ratio
the terms $R_{jj'}^{m_0}$ cancel and 
the form of the coupling Hamiltonian $H_{sb}$ is no longer relevant.
One gets 
$R_{jj'}/R_{j'j}\approx
\exp [\beta(\varepsilon_j-\varepsilon_{j'}+m_0\omega)]$, much as in
the time independent case, but reflecting the need for $m_0$ quanta
of the driving field. 

In general, the states are mixed by the driving field,
giving rates $\sum_mR_{jj'}^m$ where more than one value of $m$ contributes
significantly. Consequently the above ratio
$R_{jj'}/R_{j'j}$ will no longer depend just on a single quasienergy 
difference. 
%In fact, even for zero temperature this ratio is not simply zero or infinite
%as it is in the static case.
In order to find the stationary weights $p_j\equiv \rho_{jj}$ 
one always has to solve Eq.~(\ref{diagonal}).
This solution will in general not fulfill the detailed balance equations.
Even for the simple situation above, where the
direct and inverse rates for states $j$ and $j'$ have a simple 
Boltzmann factor ratio, we can not then conclude that the corresponding
probabilities $p_j$ and $p_{j'}$ have the same ratio. As in the case 
of a time independent
Hamiltonian, the stationary solution of Eq.~(\ref{diagonal}) 
will not depend on the overall amplitude $\gamma$ of the coupling Hamiltonian 
(if $\gamma$ is sufficiently small for lowest order perturbation theory to 
be valid). In contrast to the time-independent case, 
however, the solution {\it will} depend on the 
rates $R_{jj'}^m$ and thus on the precise form of the coupling 
Hamiltonian $H_{sb}$, as noted in~\cite{kohn}.

\section{Isolated weakly avoided crossing}

As a function of the strength of the driving field $\lambda$  the quasienergy
spectrum of  Floquet systems is generally pervaded by weakly avoided
crossings, near which quasienergy separations are small and the
Floquet states are rapidly varying. These features can have important
consequences for the density matrix.
We consider a finite set of Floquet states for which, in the neighborhood of 
the driving field strength $\lambda$ being studied, there is a single weakly
avoided crossing, with small quasienergy 
splitting $\Delta$ at the center of the crossing 
at $\lambda_0$ near $\lambda$. 
At the same time all other quasienergy separations at $\lambda$
are assumed to be large compared to the appropriate measure of the
coupling strength to the bath.  Then only the two Floquet states involved in the
weakly avoided crossing are sensitive functions of $\lambda$ over the small
crossing region.  
For very weak coupling to the bath compared
with $\Delta$ we know from the results of Sec. IIB that $\rho$ is diagonal in the
basis of those Floquet states.  In contrast to the time independent case, where
the density matrix elements depend only on the state energies,  here it will change 
rapidly as the Floquet states do, not only within the subspace of the states of 
the avoided crossing but,
as we will show, 
in general throughout the rest of the density matrix, as well.
In the opposite limit, when the system-bath interaction effectively broadens
those two states by much more than $\Delta$, we will show that there is little influence
from the crossing on the density matrix (in a $\lambda$-independent basis)
or on the statistical properties of the system
determined therefrom.   This will allow us in the next section to extend our analysis 
to many driven systems for which
the  quasienergy spectrum is dense, as is often the case in practice.

We review briefly the description of the states and
quasienergies in the neighborhood of such a weakly avoided
crossing, introducing the notation which will be used later.
We assume that within a suitably small neighborhood of driving
strengths around $\lambda_0$ the time-periodic parts 
$|u_1(t)\rangle$ and $|u_2(t)\rangle$ of the unperturbed ``diabatic" states 
(i.e., if there were no avoided crossing) do not depend 
significantly on $\lambda$,
and that the unperturbed quasienergies are locally linear in $\lambda$ 
with slopes $\sigma_1$ and $\sigma_2$, respectively.
Then, the quasienergies and the time-periodic part of the Floquet states 
can be well described by the results of diagonalizing a $2 \times 2$ 
Hamiltonian matrix (the Floquet analog of nearly degenerate perturbation
theory, using the extension of Rayleigh-Schr\"odinger perturbation 
theory developed by Samb\'e \cite{sambe}).
The quasienergies are given by
\begin{equation}
\label{eigenvalues}
\varepsilon_{a,b} = \varepsilon_0 
    + \frac{(\sigma_1 + \sigma_2)}{2} (\lambda-\lambda_0)
			\pm \frac{1}{2}\sqrt{(\sigma_1-\sigma_2)^2(\lambda-\lambda_0)^2 +
			\Delta^2}
, 
\end{equation}
where the indices $a,b$ refer to the two ``adiabatic" Floquet states continuous in
$\lambda$, and $\Delta$ is the minimum quasienergy splitting, at 
the center of the crossing, where $\lambda = \lambda_0$. This is 
shown schematically in Fig. 1.
The adiabatic states $a$ and $b$ 
exchange their character when $\lambda$ is varied,
while by assumption the diabatic states $1$ and $2$ remain unchanged.

We define the dimensionless distance $d$ from the avoided crossing 
at $\lambda = \lambda_0$ as
\begin{equation}
\label{smalld}
d \equiv (\lambda-\lambda_0)(\sigma_1-\sigma_2)/ \Delta
,
\end{equation}
the ratio of the quasienergy separation of the unperturbed
states at $\lambda$ to the gap $\Delta$. 
The central region of the avoided crossing is thus given by $|d|<1$. 

\begin{figure}[hbt]
\begin{center}
\epsfxsize=7.6cm
\leavevmode
\epsffile{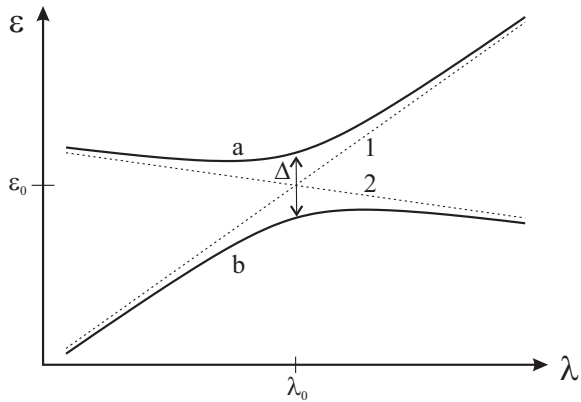}
\end{center}
\caption{Behavior of quasienergies around an avoided crossing of size $\Delta$. 
The dashed lines show diabatic (unmixed) states $1,2$ and the solid curves
the adiabatic Floquet states $a,b$.
}
\end{figure}

\subsection{Diabatic basis and effective rate $R^{\text{ac}}$}

Over the narrow range of the weakly avoided crossing the diabatic states
$|u_1(t)\rangle$ and $|u_2(t)\rangle$, as well as the time periodic parts of all other 
Floquet states, are approximately independent of $\lambda$, so they constitute
a natural fixed basis of time periodic functions to use in studying this region.
In Appendix~\ref{derivationRac} we derive the set of rate
equations for the reduced density matrix elements 
$\rho_{ij}$ in this diabatic basis.
Quite strikingly, under a number of plausible assumptions,
they allow for a separate determination of the diagonal and 
off-diagonal elements.  The former are simply given by
\begin{equation}
\label{ac3}
0 =\rho_{ii} \sum_j \bar{R}_{ij}
	- \sum_j\rho_{jj} \bar{R}_{ji} ,
\end{equation}
where the rates designated with a bar,
$\bar{R}_{ij}$, are defined in the diabatic basis in analogy to Eq.~(\ref{realrates}), 
except for
\begin{equation}
\label{defRAC}
\bar{R}_{12} = \bar{R}_{21} = R^{\text{ac}} \equiv 
\frac{\Gamma }{4d^2 + ( \Gamma /\Delta )^2} ,
\end{equation}
with the superscript on $R^{\text{ac}}$ a reminder that this is an effective rate 
associated with the small quasienergy separation near the avoided crossing (ac).  
Here we have introduced the symbol
\begin{equation}
\label{defr}
\Gamma  \equiv 
{\sum_k}' \bigl( \bar R_{1k} + \bar R_{2k} \bigr) 
+ \bar R_{11} + \bar R_{22}- 2\bar R_{11;22}
,
\end{equation}
where the prime on the summation denotes the restriction $k \neq 1,2$.
(Note that the sum of the last 3 terms is positive, in spite of
the minus sign; by definition they can be written 
$2\pi\gamma^2\sum_m[A_{11}(m)-A_{22}(m)]^2g(-m\omega)$.)
The rate $\Gamma$ will turn out to be the relevant coupling to the 
heat bath, specific to the crossing of states 1 and 2,
that will be the measure of whether that coupling can
be considered weak or strong relative to the quasienergy splitting of those states.

These equations~(\ref{ac3}) for the diagonal density matrix elements~$\rho_{ij}$
in the diabatic basis have exactly the formal structure of  
the corresponding equations in the limit of bath coupling small compared to all
quasienergy splittings, Eq.~(\ref{diagonal}).  
The results differ from that simpler situation in three essential
ways: 

 (i) The effective rates $R^{\text{ac}}$ between the two special
states $1$ and $2$ are not simply the second order rates imposed 
directly by the bath.  They reflect the mixing in a more
complex way, effectively including the driving induced coupling to all orders,
and they can be simply determined from the rate $\Gamma $ defined 
in Eq.~(\ref{defr})
and the properties of the avoided crossing, namely the distance $d$  
and the minimal quasienergy splitting $\Delta$.

(ii) The dependence on the parameter $\lambda$,
which in the original adiabatic basis strongly affects a great number of 
rates, appears in the diabatic basis only
in the single rate $R^{\text{ac}}$  (see Eq.~(\ref{defRAC}), where
the distance from the avoided crossing, $d$, is linearly related to $\lambda$).

(iii) These are not the {\it only} density matrix elements; 
there exist also two non-vanishing off-diagonal elements $\rho_{12}$ 
and $\rho_{21}=\rho_{12}^*$ in this diabatic basis, given by
\begin{equation}
\rho_{12} = \frac {\rho_{11}-\rho_{22}}{2d-i\Gamma/\Delta} .
\label{rho12}
\end{equation}
Their importance can be assessed by a
comparison of the absolute value of $\rho_{12}$ to the sum of the positive and 
real values $\rho_{11}$ and $\rho_{22}$: 
\begin{equation}
\frac{|\rho_{12}|}{\rho_{11}+\rho_{22}} 
= \frac{|\rho_{11}-\rho_{22}|}{\rho_{11}+\rho_{22}} 
\frac{1}{\sqrt{4d^2 + (\Gamma /\Delta)^2}}
\leq \frac{\Delta}{\Gamma } ,
\end{equation}
which is negligible in the limit $\Gamma  \gg \Delta$,
(that is, where the coupling to the heat bath
is stronger than the minimum quasienergy separation of the Floquet states).
This demonstrates that in this limit the density matrix in this two-dimensional
subspace is approximately diagonal in the diabatic basis.
In the opposite limit, the density matrix in the subspace
is nearly diagonal in the adiabatic Floquet basis.  The appropriate
basis to diagonalize $\rho$  is given in general in 
Appendix~\ref{diagonalizingmatrix}.

Although we have no general analytic solution of the equations~(\ref{ac3})
for the diagonal density matrix elements,
we can say a good deal about the consequences of a weakly avoided crossing,
as the relevance of the effective rate $R^{\text{ac}}$ depends crucially
on the ratio of the bath-induced rate $\Gamma$ to the
size of the avoided crossing $\Delta$.

\subsubsection{$\Gamma \ll \Delta$}

In this limit (which is just the case discussed in Sec. IIB above)
the rate $\Gamma$ characterizing the coupling to the heat bath
is even smaller than the quasienergy splitting at the weakly avoided crossing.
Consequently, the effective rate $R^{\text{ac}}$, Eq.~(\ref{defr}), is much larger than
$\Gamma$ in the center of that crossing,
\begin{equation}
	R^{\text{ac}}(d=0) \gg \Gamma
	.
\end{equation}
The immediate consequence of this large effective rate is that the
diagonal density matrix elements $\rho_{11} \approx\rho_{22}$ become
equal there, which may drastically change all other weights $\rho_{jj}$,
coupled directly or indirectly via Eq.~(\ref{ac3}) to 
$\rho_{11}$ or $\rho_{22}$, relative to the corresponding values just outside
the avoided crossing.
An explicit example demonstrating such a drastic change for the entire density matrix
is studied in Sec.~\ref{simpleexample}.

The off-diagonal matrix element $\rho_{12}$ is not necessarily small, though, 
since it is given by the product of one large and one small factor; see 
Eq.~(\ref{rho12}). In fact, as discussed in Appendix~\ref{diagonalizingmatrix}, the off-diagonal
element is negligible in the adiabatic basis only.

\subsubsection{$\Gamma \gg \Delta$}
\label{IIIA2}

In this limit the rate $\Gamma$ induced by the coupling to the heat bath
is stronger than the size of the weakly avoided crossing.
Consequently, the effective rate $R^{\text{ac}}$, Eq.~(\ref{defr}), is much smaller than
$\Gamma$ throughout the whole range of the avoided crossing,
\begin{equation}
	R^{\text{ac}} \ll \Gamma
	.
\end{equation}
This suggests that the effective rate $R^{\text{ac}}$ due to the avoided crossing 
has negligible influence compared to the other rates affecting states 1 and 2.
Therefore the entire density matrix is not affected by the existence
of the weakly avoided crossing.
The off-diagonal matrix element $\rho_{12}$ is negligible in this limit.

There is an exception to the above conclusion if the two sums
$\sum_k'  \bar R_{1k} $ and
$\sum_k'  \bar R_{2k} $
contributing to $\Gamma$ have vastly different magnitudes.
This is possible, as we are explicitly
looking at a very weakly avoided crossing, and this typically arises
from such dissimilar pairs of states, such as those arising from
undriven states one of which is of low and the other of very high
energy. In what follows we will take state 2 to be the one
with higher average energy $\langle H_s\rangle$.
There are two cases: 
\begin{itemize}

\item
$\sum_k' \bar R_{2k} \ll R^{\text{ac}} \ll \sum_k' \bar R_{1k}$: 
This case may occur if the heat bath is dominated by phonons (or by other
bosons) with an upper energy limit $\omega_D$, small compared to the 
separation in average energy of the diabatic state $2$ and its neighbors, so that state
$2$ will decay primarily to state $1$,
with flux $\rho_{22}R^{\text{ac}}$. Similarly, state $2$
is fed primarily by the direct process from state $1$ with flux 
$\rho_{11} R^{\text{ac}}$ (in the extreme limit of all $\bar R_{j2}=0$, {\it only}
by this process).  Since stationarity
demands that the incoming and the outgoing flow for state 2 are equal, 
one finds $\rho_{11} \approx\rho_{22}$, with corresponding large changes
in other state occupations $\rho_{jj}$. 
This modification of
the density matrix occurs for all values of $d$ and $\Gamma/\Delta$
for which the presumed rate inequalities hold.  Values of those parameters 
can be significantly larger than unity, but eventually large enough values 
will reduce $R^{\text{ac}}$ to the point where the lower inequality
is violated.

\item
$\sum_k' \bar R_{1k} \ll R^{\text{ac}} \ll \sum_k' \bar R_{2k}$: 
This case may occur when the heat bath is dominated by the
electromagnetic radiation field, with photons of arbitrarily high energy
and large density of states.  Then state $2$, essentially a high energy
undriven state, will decay rapidly. 
By precisely the same argument used in the previous case, 
$\rho_{11} \approx\rho_{22}$, and states now strongly connected 
to $2$ acquire substantial occupancy.
Again, the ranges of $d$ and $\Gamma/\Delta$ over which this
situation holds can be large.
\end{itemize}

\subsection{A simple example}
\label{simpleexample}

It is instructive to obtain specific results for a simple, albeit somewhat
artificial, example of a system.
We assume first that for Floquet states labeled by some index, $n=1,2,\dots N$, 
the thermal
bath leads to  coupling only of neighboring states --- i.e.,
the only non-zero rates are of the form $R_{n, n \pm 1}$, as shown in Fig.~2,
and that we are in the limit of coupling to the bath weak with respect to all
quasienergy differences.
In this special situation the lowest state $1$ has to be in
detailed balance with state $2$ and from this it follows that 
all states are in detailed balance.
If we make the specific choice $R_{n,n-1} = R_d$ and $R_{n,n+1} = R_u$, each
independent of $n$, and denote the ratio $R_u/R_d \equiv \gamma < 1$,
then the occupations $p_n \equiv\rho_{nn} = (1-\gamma) \gamma^{n-1}/(1-\gamma^N)$
fall off exponentially with the index $n$.  We assume $N$ to be sufficiently large that
we can neglect (for algebraic simplicity) the factor $\gamma^N$ below.

Now we want to study how these occupations change if there is a single
weakly avoided crossing of states $n_1$ and $n_2$ 
(in the previous section these were the states called $1$ and $2$).
This introduces the additional rate $R^{\text{ac}}$.
It simplifies the algebra to define the fluxes
between states $n$ and $n-1$,
\begin{equation}
\label{deffn}
F_{n,n-1} \equiv p_n R_{n,n-1} - p_{n-1} R_{n-1,n} ,
\end{equation} 
in terms of which we can rewrite the basic equations derived above
for the diagonal density matrix elements, (\ref{ac3}), as
\begin{eqnarray}
\label{srf1}
(p_{n_1} -p_{n_2}) R^{\text{ac}} &=& F_{n_2,n_2-1} - F_{n_2+1,n_2} \\
\label{srf2}
(p_{n_1} -p_{n_2}) R^{\text{ac}} &=& F_{n_1+1,n_1} - F_{n_1,n_1-1} \\
\label{srf3}
0 &=& F_{n+1,n} - F_{n,n-1} \qquad \text{for} \qquad n \neq n_1,n_2 
\quad \text{and} \quad  n \neq 1 \\
\label{srf4}
0 &=& F_{2,1} .
\end{eqnarray}
Their solution is
\begin{equation}
F_{n,n-1} = \left\{ \begin{array}{ccl}
                   0 & ; & n \leq n_1 \\
                   F & ; & n_1 < n \leq n_2 \\
                   0 & ; & n > n_2  
                  \end{array} \right. ,
\label{feqn}
\end{equation}
with 
\begin{equation}
F = (p_{n_1} - p_{n_2}) R^{\text{ac}} ,
\label{fdef}
\end{equation}
being the flux from $n_1$ to $n_2$ due to the avoided crossing.
This result is illustrated 
in Fig.~2. The states above $n_2$ and the states below $n_1$ are still in 
detailed balance --- i.e. $F_{n,n-1}=0$, as without the avoided crossing. 
The states from $n_1$ to $n_2$, however, are no longer in detailed balance and
there is a stationary flux $F$ connecting them, due to the avoided crossing.

\begin{figure}[hbt]
\begin{center}
\epsfxsize=5.0cm
\leavevmode
\epsffile{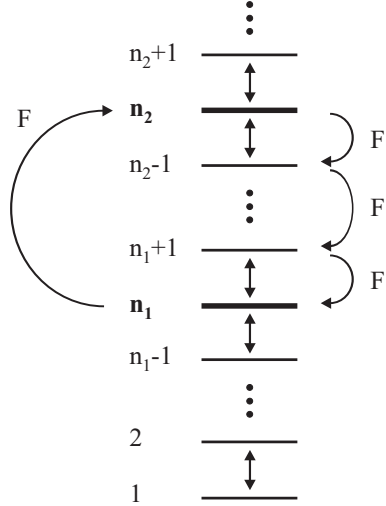}
\end{center}
\caption{Simple example with a thermal bath coupling neighboring states only (double
headed arrows) and flux $F$ introduced by the effective rate $R^{\text{ac}}$
due to an avoided crossing of states $n_1$ and $n_2$.
}
\end{figure}

For the occupations $p_n$ we find
\begin{equation}
p_n = \left\{ \begin{array}{lcl}
             p_1 \gamma^{n-1} & ; & n \leq n_1 \\ \\
             p_{n_1} \frac{r + \gamma^{n-n_1}}{r+1} & ; & n_1 < n \leq n_2 \\ \\
             p_{n_2} \gamma^{n-n_2} & ; & n > n_2  
            \end{array} \right. ,
\label{peqn}
\end{equation}
with 
\begin{equation}
r \equiv \frac{ R^{\text{ac}}}{R_d}
\frac{1-\gamma^{n_2-n_1}}{1-\gamma} ,
\end{equation}
and $p_1$ determined by the normalization $\sum_n p_n =1$.

\begin{figure}[hbt]
\begin{center}
\epsfxsize=7.6cm
\leavevmode
\epsffile{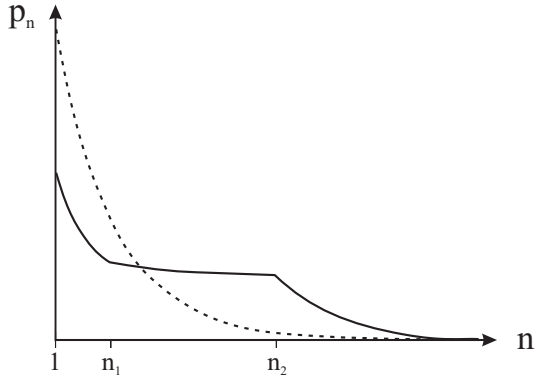}
\end{center}
\caption{Dramatic change in occupations $p_n$ with (solid curve) and 
without (dashed curve) avoided crossing.
}
\end{figure}

The parameter $r$ measures the ratio of the effective rate $R^{\text{ac}}$
describing the direct connection of the state $n_2$ to the state with 
which it is crossing, namely $n_1$, relative to  
its rate of loss $R_d$ to the state directly below it, times a factor of order one.
There are three distinct interesting ranges of $r$ in terms of the
behavior of the density matrix elements:

(i) When $r\gg 1$ we have $p_{n_2} \approx p_{n_1}$ and almost constant
weights between $n_1$ and $n_2$.
The weights of states above $n_2$ are much larger 
than they would be without the avoided crossing, due to
the drastically increased weight $p_{n_2}$, which is being fed
directly by $n_1$.

(ii) For $ \gamma^{n_2-n_1}\ll r \ll 1 $ we have $p_{n_2} \approx p_{n_1} r$, 
still a much
higher occupancy than in the absence of the avoided
crossing. 
Again the weights of states above $n_2$ are much larger 
than they would be without the avoided crossing.
This case, with $r$ only
slightly less than 1, is sketched in Fig. 3.

(iii) If $r$ decreases still further,  $r \ll \gamma^{n_2-n_1}$,  then 
$p_n \approx p_1 \gamma^{n-1}$, which is the
result without the avoided crossing.
Equilibration with the bath will be completely dominated by the
transition processes along the ladder of states.

This model admittedly has some special features which make it
soluble.  For instance, there is detailed balance in the absence of
an avoided crossing.  But it illustrates the dramatic change in all 
occupations due to a single avoided crossing.

\section{Limit of infinite dimensional Hilbert space}

As we have already noted, there are some interesting Floquet systems, such as finite spin 
systems, which are limited to a space spanned by a finite number of basis states. Their 
statistical mechanics can then be calculated (numerically) from the equations and analysis 
of the previous sections of this paper (and we will not be concerned with them further here). 
However, many others, like a particle confined by a potential well, have eigenstates spanning 
infinite Hilbert spaces, with a quasienergy spectrum which becomes dense and Floquet states 
which change rapidly as the basis size N increases, so that the above results do not readily 
apply. In fact, our results also permit the calculation to any desired accuracy of many of 
these, as well, by appropriate truncation of the basis set, as we now show. 

We limit ourselves to systems with an unbounded discrete spectrum of {\it undriven} energy 
eigenstates.  We take the truncated Hilbert space spanned by the $N$ lowest energy
eigenstates,  the usual choice made in numerical
calculations.  Within this approximation the Floquet states of the isolated system
(at any fixed time)
are confined to that finite dimensional space.  For the undriven system the statistical
(Boltzmann) weights of the states neglected by the truncation can be made arbitrarily 
small by a suitably large choice of the energy cutoff.  We now increase the 
Hilbert space accessible to the driven system by the addition of the $(N+1)$st undriven energy
eigenstate.   With increasing $\lambda$, as its quasienergy nears that of one of 
the original $N$ Floquet states, it will mix significantly with that state.  But
we have shown above (section~\ref{IIIA2}) that, if the corresponding avoided crossing is
sufficiently weak, $\Delta \ll \Gamma$, then the reduced density matrix is approximately diagonal
in the unmixed diabatic states, and the addition of the new state to the basis set
remains of little statistical significance.   Therefore, if we can demonstrate that
all of the states neglected by the truncation of high energy undriven eigenstates
mix with those retained only through such weakly avoided crossings, then we can choose the
energy cutoff of the basis to obtain statistical properties to any desired accuracy.

Indeed, this desired behavior of the quasienergy of a Floquet state arising from a high 
energy undriven eigenstate has been seen in numerical calculations~\cite{holthaus}.  
Such a state is found to mix
with the Floquet states that arise from low energy undriven eigenstates,
as $\lambda$ increases, only near weakly avoided crossings.  In the appendix 
of Ref.~\cite{hkk} we have analytically demonstrated the effect
explicitly, with an exponential decrease with basis size $N$ of newly introduced quasienergy
gaps, for systems whose undriven energy eigenvalues above
some level exhibit increasing spacings between successive levels (such
as a particle confined by a one-dimensional potential which increases faster than 
harmonically).  This can be rationalized in terms of a perturbation expansion:  
an increasingly large number
of driving field quanta is needed to connect to higher lying states, which are
modified only in increasingly higher order perturbation theory.  More generally,
numerical studies~\cite{bdh} suggest that once the first $N$ undriven states have
been included in the basis, the Floquet states arising from the first $N'$ of
these, for suitably large $N-N'$, are affected by further extension of the basis set 
only by weakly avoided crossings.  Again, perturbation theory suggests 
that this is a reflection of the very different spatial behavior
of these states.

For a class of periodically driven systems in an infinite Hilbert space we earlier found~\cite{hkk} 
many anomalies due to the infinite number of ever more weakly avoided crossings.
The present results show that these anomalies, and the non-convergence of the
Floquet states with increasing basis size, lead to negligible  
consequences for the reduced density operator, or for the statistical
properties calculated therefrom, when the system is in {\it fixed} weak contact 
with a thermal bath.

In practice, one has to determine the quasienergy
spectrum for increasing basis size $N$ until (i) newly appearing
avoided crossings are small enough, $\Delta \ll \Gamma$, and
(ii) the statistical (Boltzmann) weight of the newly introduced states,
plus that of all neglected states,  is small enough to achieve 
the desired accuracy. 
%(Note, that
%the first criterion, $\Delta \ll \Gamma$, in principle has to be
%checked individually for every avoided crossing, as $\Gamma$,
%Eq.~(\ref{defr}), is defined in terms of the matrix elements between
%the two states involved in a specific crossing.)  
Once the
basis size $N$ is appropriately chosen, one can
use Eq.~(\ref{rate_equation_general}) to determine the density
matrix elements and calculate statistical properties.

It is important to recognize that, because of the requirement $\Delta \ll \Gamma$
for avoided crossings with all neglected basis states, the minimum truncation size $N$
depends in general on the
{\it magnitude}, as well as the form, of the system-bath coupling, since
both enter the determination of $\Gamma$.  Indeed, the full statistical properties 
of the Floquet system will depend on that magnitude 
(we have already emphasized in Sec. IIB the general dependence on
the {\it form} of the coupling), even if the coupling is very weak: 
For a given magnitude we choose $N$ to achieve a desired accuracy.
We now reduce the magnitude (keeping the form fixed) 
sufficiently that $\Gamma$ becomes
less than one of the gaps $\Delta$ associated with an
avoided crossing of a state neglected by the truncation with a state
that has been retained.  Then that avoided crossing becomes no longer 
irrelevant.   It leads to a large rate
$R^{\text{ac}}$, Eq.~(\ref{defRAC}), that may cause quite different
statistical weights, as was explicitly shown for the simple example
in Sec.~\ref{simpleexample}.  We can understand this
as the appearance of a new, high order, near resonance which is
no longer disrupted by the bath interaction, resulting in the appreciable
occupation of the upper (higher average
energy) level of that pair, and modifying the statistical weight 
of the states to which it is connected.

This is emphatically not to suggest the lack of a well-defined statistical
behavior of a Floquet system with given coupling to a bath.  Indeed, the above 
procedure describes how that is to be calculated to any desired accuracy.
However, the behavior does depend on the specific magnitude and form
of that coupling, in sharp contrast to the situation for conventional 
time-independent statistical mechanics.

\section{Conclusion}

We have derived the general equations~(\ref{rate_equation_general}) 
for the reduced density matrix, in the stationary state, of Floquet 
systems with weak but 
{\it finite} coupling to a heat bath, previously discussed~\cite{blumel,kohler,breuer,kohn}
only for situations where all quasienergy spacings were large compared with the coupling
to the bath (and therefore only for systems restricted to a finite number of states).
New and interesting results give a prescription for practical calculation of
the steady state statistical properties of many realistic Floquet systems, including
ones with dense quasienergy spectra.

If the coupling to the bath $\Gamma$ is restricted only to be
small compared to the driving frequency
$\omega$  and the inverse bath relaxation time $1/\tau_c$, the
diagonal and off-diagonal matrix elements are approximately time-independent
in the basis of the time-periodic parts of the Floquet states. 
In the special case of 
a Hilbert space truncated to finite dimensions and the coupling weak
compared to all quasienergy separations, then
the equations simplify and are equivalent to those discussed
in~\cite{kohn} and~\cite{breuer}. The density matrix is diagonal,
but the values of the diagonal elements (occupation probabilities
of the Floquet states) depend on the detailed form (though not on
the magnitude, as long as it is sufficiently weak) of the coupling
of the system to the bath.

Weakly avoided crossings, where quasienergies become nearly
degenerate, are a pervasive feature of Floquet systems.
Therefore, we have analyzed in detail the general density matrix
equations in such regions. The rate equations for diagonal density 
matrix elements in the diabatic basis, Eq.~(\ref{ac3}),
are shown to be identical to those well away
from the crossing and with arbitrarily weak coupling to the heat bath,
except that the avoided crossing introduces a new effective rate 
$R^{\text{ac}}$,
defined in Eq.~(\ref{defRAC}), between the  diabatic
states 1 and 2 of the crossing.
This rate depends on the effective coupling $\Gamma$ of the states 1 and 2
to the heat bath, the distance $d$ from
the avoided crossing, and the relative strength of the coupling to the 
heat bath, $\Gamma/\Delta$.
In addition, significant off-diagonal density matrix elements $\rho_{12}$ 
and $\rho_{21}$  may appear (see Eq. (\ref{rho12})).

The density operator changes drastically with $\lambda$
in the neighborhood of an avoided crossing, $d \lesssim 1$, if also the bath
coupling is weak enough, $\Gamma  \lesssim \Delta$. 
Quite importantly, significant changes occur in the weights not only of the 
states mixed in the crossing, but in those of many other states, as well.
At the center of the crossing the two states of the crossing 
pair acquire equal weight, in general a very different
relative occupation from that outside the 
crossing region, and this consequently modifies significantly the
weights of the network of states with which they interact, directly
and indirectly, as described by the rate equations. This is exhibited 
explicitly for the simple example worked out in Sec. IIIB,
where the occupations of all states between those labeled $n_1$ and 
$n_2$ are strongly modified in this regime. On the other hand, if the
coupling to the bath is relatively large, $\Gamma \gg  \Delta$, then
physical intuition suggests that
the interaction with the bath effectively broadens the quasienergy
dispersion curves by more than the gap itself, 
such that the avoided crossing should have no consequences
on the statistical properties of the system.
We have shown that this is indeed the case,
namely that the density matrix (in a $\lambda$-independent basis) remains 
virtually unaffected.

Having studied in detail Floquet systems spanned by a finite
basis, we have looked at the consequences when the size of the
basis set increases without limit, as for a particle confined by
an unbounded potential and driven periodically.  We have concluded that,
in many cases,
the important anomalies of such systems, including the appearance 
of an infinite number of weakly 
avoided crossings
and the non-convergence of the
Floquet states with increasing basis size,
play no important role for the physical properties
of the system in contact with a 
heat bath, confirming our physical intuition. One can safely
employ, in numerical calculations, a finite suitably truncated 
basis set for such a system.  However, a special feature of such a 
Floquet system, which has a dense quasienergy spectrum, is that
its statistical properties depend in general on both the magnitude and the
overall strength of the coupling between the system and the bath.

These calculations have been based
on a perfectly periodic driving field.  We recognize that effects that
we have neglected, notably fluctuations in amplitude and frequency
of the driving field, are similarly expected to wash out the effects of 
sufficiently weakly avoided crossings.  Those effects will be examined
in a future publication.

\section*{Acknowledgement}
We thank S. Kohler and W. Wustmann for helpful discussions.
RK thanks the Kavli Institute for Theoretical Physics, Santa Barbara 
for hospitality.  DH is grateful for the partial support of the Edward A. Dickson Emeriti
Professorship and NSF Grant PHY05-51164.

\appendix
\section{Validity of lowest order Born approximation}
\label{4thorder}
In the still exact Eq.~(\ref{exactint}) we have made the lowest order
Born approximation by using the factorization
Eq.~(\ref{born}) of the total density matrix directly on the right hand side 
of that equation.  In order to check the validity
of this approximation, we analyze in this appendix
the fourth order correction arising from the factorization of the total 
density matrix only in the first iterate of Eq.~(\ref{exactint}), namely
\begin{eqnarray}
&&  + \int_0^t dt' \int_0^{t'} dt'' \int_0^{t''} dt''' 
{\rm Tr_b} \left[\tilde H_{sb}(t),\left[\tilde H_{sb}(t'),
\left[\tilde H_{sb}(t''),\left[\tilde H_{sb}(t'''),\tilde\rho(t''') \otimes \rho_b \right]\right]  
\right]\right]
\\ \nonumber
&& - \int_0^t dt' \int_0^{t'} dt'' \int_0^{t''} dt''' 
{\rm Tr_b} \left[\tilde H_{sb}(t),\left[\tilde H_{sb}(t'), {\rm Tr_b} \left(
  \left[\tilde H_{sb}(t''),\left[\tilde H_{sb}(t'''),\tilde\rho(t''') \otimes \rho_b \right]\right]
  \right) \otimes \rho_b \right]\right]
\end{eqnarray}
Each of these two terms has the structure of three nested time integrals, with a sum
of integrands from the nested commutators (the terms differing only by the order
of the factors), each of which is a product of 4 system 
operators and a 
fourth order correlation function of the bath.  A representative such bath correlation
function is
\begin{eqnarray}
&& \langle  \tilde B(t) \tilde B(t') \tilde B(t'') \tilde B(t''')  \rangle
	-	\langle \tilde B(t) \tilde B(t') \rangle \langle \tilde B(t'') \tilde B(t''') \rangle
	\\&&=
	\langle \tilde B(t) \tilde B(t'') \rangle \langle \tilde B(t') \tilde B(t''') \rangle
	+ \langle \tilde B(t) \tilde B(t''') \rangle \langle \tilde B(t') \tilde B(t'') \rangle
	\\&&=
	 G(t-t'') G(t'-t''') + G(t-t''') G(t'-t'')	
	,
	\label{4thordercorrelation}
\end{eqnarray}
where the equality of (A3) to (A2) makes use of the generalized Wick theorem \cite{fetter}.
The final pair of correlation function products (A4) has the important property
that in each of the two terms the time intervals overlap. 
(The only combination where this is not the case is explicitly subtracted in (A2).)
Thus all three times appearing in the integrals, $t'$, $t''$, and $t'''$, 
must be close to $t$ in order to give
an important contribution, since $G(\tau)$ 
is significant only for $\tau < \tau_c$.
This same condition holds for {\it all} contributions to the fourth order term.
Each such term has an upper limit given by
\begin{equation}
	\Gamma_4 \equiv \gamma^4 \tau_c^3 \langle A^4 \rangle \langle B^2 \rangle ^2,
\end{equation}
which compared to the magnitude of the second order rate $\Gamma_2$
is small if 
the weak coupling condition, $\tau_c \Gamma_2 \ll 1$, 
Eq.~(\ref{bornmarkovcondition}),
is fulfilled,
namely that during the bath correlation time $\tau_c$ 
the impact of the second order terms is small; Eq.~(\ref{bornmarkovcondition})
is itself sufficient to ensure the smallness of higher order corrections.

A more restrictive condition,
$\Gamma_2 < \varepsilon_{ij}$ for all $i,j$,
was imposed in Ref.~\cite{kohler},
namely that the system-bath coupling must be
smaller than the smallest quasienergy splitting.
In contrast, the above estimate of the fourth order term
shows that larger system-bath couplings with $\Gamma_2 > \varepsilon_{ij}$
can equally well be treated with the present approach,
as long as the weak coupling condition, Eq.~(\ref{bornmarkovcondition}), 
is fulfilled. 

\section{Validity of time averaging of the rates}
\label{validitytimeaverage}

The replacement of rates by their averages over a driving period 
(see Eq. (\ref{averagerates}) and the sentence preceding it) 
relies on the assumption that the density matrix elements
do not vary substantially over that period, $T=2\pi/\omega$.
At first glance this seems to be wrong as, e.g., an off-diagonal density matrix
$\rho_{ij}(t)$ with an associated quasienergy difference
$\varepsilon_{ij}$, that can be as large as
$\omega$, would have in the absence of the coupling to the
heat bath a substantial variation over the time $T$.
But it will turn out that such off-diagonal matrix elements (corresponding
to large quasienergy differences) are in fact small
at large times.

We use the observation that the 
set of differential equations (\ref{time_rate_equation_general}) is
linearly coupled and has time-periodic coefficients $R_{lj;ki}(t)$ 
with period $T$.
Thus one can apply the Floquet theorem, which requires the
general solution to be of the form
\begin{equation}
\rho_{ij}(t) = \sum_n a_n e^{\lambda_n t} r_{ij;n}(t)
,
\end{equation}
with time-periodic functions $r_{ij;n}(t+T)=r_{ij;n}(t)$, and
$e^{\lambda_n T}$ an eigenvalue of the
time-evolution operator for all density matrix elements over one period.
(The $\lambda_n$ are assumed here to be non-degenerate.)
This solution is in general not periodic in time.

In the following we are interested in the solution for
large times.
All solutions with $Re(\lambda_n) < 0$ decay to zero in this limit
and are of no further interest.
From the general properties of a density matrix, $0 \le \rho_{ii} \le 1$
and $|\rho_{ij}|^2 \le \rho_{ii} \rho_{jj} \le 1$,
it follows that solutions with $Re(\lambda_n) > 0$ are impossible.
So we are left with discussing solutions with $Re(\lambda_n) = 0$.
For the trace of the density matrix to be 1, i.e. 
time-independent and nonzero, there must be a solution $\lambda_0=0$.
Thus we have a stationary solution for the density matrix elements
\begin{equation}
\rho_{ij}(t) = r_{ij;0}(t)
,
\end{equation}
which is time-periodic, with at least some diagonal components which have
non-zero time averages. (We cannot exclude that in addition to  
this, there is a traceless
quasiperiodic solution from pairs of purely imaginary $\lambda_n$.
As its time average would be zero, we ignore it in the following.)

We now expand this stationary time-periodic solution in
Fourier coefficients
\begin{equation}
\rho_{ij}(t) = \sum_K \rho_{ij}(K) e^{iK\omega t}
,
\end{equation}
so that the above differential equations  
(\ref{time_rate_equation_general})
reduce,
for each of the Fourier components, to
\begin{eqnarray}
\label{rhoK}
-i (\varepsilon_{ij} + K\omega) \rho_{ij}(K) 
&=&
\frac{1}{2} \sum_{k,l,M} \Bigl[
    \rho_{lj}(K-M) R_{ik;lk}(M) + \rho_{il}(K-M) R_{jk;lk}^*(-M)\\ \nonumber
   &-&\rho_{kl}(K-M) (R_{lj;ki}(M) + R_{ki;lj}^*(-M)) \Bigr]
\end{eqnarray} 
The right hand side is second order in the small system-bath coupling
constant $\gamma$, because each rate factor (\ref{Km_complexrates}) is.   
Although we are explicitly concerned with
couplings which are not arbitrarily weak with respect to the
smallest quasienergy differences in the problem, we do want to consider,
as in the conventional static case, coupling which is weak on
higher energy scales, such as the driving frequency $\omega$.
Then for non-zero values of $K$ the necessary smallness of the 
left hand side of (\ref{rhoK}) must be achieved by the smallness 
of $\rho_{ij}(K)$, and it is consistent to neglect
all such components of the density matrix in the solution; we will set
\begin{equation}
\label{kneq0}
\rho_{ij}(K \ne 0) = 0 ,
\end{equation}
and we will simply write $\rho_{ij}$ for $\rho_{ij}(K=0)$, which is
by definition time independent. (We point out that this implies a time-periodic
density {\it operator} for large times, as one would expect physically.)
This time independence of $\rho_{ij}$ at large times then implies in (\ref{rhoK}) that
only $M=0$ terms appear --- i.e., that the complex rates $R$ can be replaced by
their time averages $R(M=0)$.
(Note that for a quasienergy difference $\varepsilon_{ij} \approx \pm \omega$
between a Floquet state close to the lower boundary and a state
close to the upper boundary of the chosen strip $0\leq\varepsilon <\omega$,
one would need to choose $K=\mp1$ in Eq.~(\ref{rhoK}) or choose a conveniently displaced
strip.)

\section{Derivation of the effective rate $R^{\text{ac}}$}
\label{derivationRac}

In this appendix we derive the rate equations for the reduced density
matrix elements in the diabatic basis of an isolated weakly avoided
crossing in the stationary limit.
This leads to an effective rate $R^{\text{ac}}$, which expresses
the entire $\lambda$-dependence related to an avoided crossing.

Starting from the operator equation~(\ref{our4.27}) one follows
the steps leading to Eq.~(\ref{rate_equation_general}), but now
in the diabatic basis of the avoided crossing,
which is related to the adiabatic basis by
\begin{eqnarray}
|u_a(t)\rangle &=& \alpha |u_1(t)\rangle + \beta  |u_2(t)\rangle \\
|u_b(t)\rangle &=& \beta  |u_1(t)\rangle - \alpha |u_2(t)\rangle
\end{eqnarray}
with $\alpha = \sqrt{(1 + d/\sqrt{1+d^2})/2}$,
$\beta =  \sqrt{(1 - d/\sqrt{1+d^2})/2 }$ 
and $d$, the dimensionless distance from the center of the avoided 
crossing, is defined in Eq.~(\ref{smalld}). 
This basis change alters the left hand side of Eq.~(\ref{rate_equation_general}), 
as the coherent dynamics is less conveniently
described in the diabatic basis.
The right hand side is left invariant,
but now with rates in the diabatic basis,
and can be simplified with the following assumptions:

\begin{enumerate}
\item 
We assume that the avoided crossing is sufficiently weak and isolated
within the quasienergy spectrum, such that 
we can neglect any dependence of diabatic states $1$ and $2$ on the driving amplitude 
$\lambda$, valid over a sufficiently limited range of $\lambda$ 
near $\lambda_0$.
We assume the same for all other Floquet states, such that the
diabatic basis is independent of $\lambda$ within that range.

\item
All other quasienergy separations near $\lambda = \lambda_0$
are assumed to be much larger than the splitting $\Delta$
as well as larger than the effective coupling to the bath,
such that all off-diagonal density matrix elements
can be neglected, except for $\rho_{12}$ 
and $\rho_{21}$.

\item
We define rates  $R_{ij;kl}$ in the diabatic basis
in analogy to Eqs.~(\ref{averagerates}) and 
(\ref{Km_complexrates}), where the $\lambda$ dependence
appearing in the quasienergies $\varepsilon_1$ and $\varepsilon_2$
is neglected in the neighborhood of the avoided crossing.
That is, the Fourier transformed bath correlation function 
$g(\varepsilon -m\omega )$ is taken to be constant over this very small
range of $\varepsilon$.
This is a good approximation as long as $\Delta \tau_c \ll 1$ holds,
which is the case for the weakly avoided crossings we are interested in.
Together with assumption 1 this makes the rates $R_{ij;kl}$
independent of $\lambda$.

\item
We set $A_{12}(m) = A_{21}(m) =0$ (see Eq.~(\ref{Aij})), 
as we assume that the spatial
structures of states $1$ and $2$ are very different.  We are explicitly
looking at a very weakly avoided crossing, and this typically arises
from such dissimilar pairs of states, such as those arising from
undriven states one of which is of low and the other of very high
energy. Similarly, any other state $j$
cannot be spatially similar to $1$ and $2$ simultaneously, 
leading to $A_{1j}(m) A_{2j}(m) =0$.

\item
For simplicity
we restrict the discussion to periodic drivings that are symmetric,
i.e. $H_s(t)=H_s(-t)$. This makes the rate $R_{22;11}$, which
appears in the definition of $\Gamma$ in Eq.~(\ref{defr}),  real.
\end{enumerate}

For the density matrix in the subspace of states 1 and 2 we find
\begin{eqnarray}
\label{rateEq_diab12_a}
   \Delta \text{ Im}(\rho_{12})
  &=&
    -\rho_{11} \sum_{k}R_{1k} 
    + \sum_{k}\rho_{kk} R_{k1}
  \\
   - 
   \Delta \text{ Im}(\rho_{12})
  &=&
    -\rho_{22} \sum_{k}R_{2k} 
    + \sum_{k}\rho_{kk} R_{k2}
  \\
\label{rateEq_diab12_c}   
   i \Delta \Bigl[(\rho_{11} -\rho_{22}) - 2d\rho_{12}\Bigr]
  &=&
     \rho_{12} \Gamma
\end{eqnarray}
with the symbol $\Gamma$ introduced in Eq.~(\ref{defr}).

The third equation expresses the off-diagonal element $\rho_{12}$
in terms of the diagonal elements, as explicitly written in Eq.~(\ref{rho12}).
Consequently, the left hand side of the first two equations can be
written in terms of the diagonal elements,
\begin{equation}
   \Delta \text{ Im}(\rho_{12})
   =
   R^{\text{ac}} (\rho_{11}-\rho_{22})
\end{equation}
where we use the notation $R^{\text{ac}}$ introduced in Eq.~(\ref{defRAC}).
We now redefine the effective rates $\bar R_{ij}$ in the diabatic basis, equal to
the original rates $R_{ij}$ except for $\bar R_{12}=R^{\text{ac}}$ 
and similarly $\bar R_{21}=R^{\text{ac}}$ (note $R_{12}=0$  by  
assumption 4 above).
This allows us to rewrite the first two equations in the standard rate equation form,
as if there were just diagonal density matrix elements: Eq.~(\ref{ac3}).

\section{Diagonalizing the density matrix}
\label{diagonalizingmatrix}
It is instructive to ask in which basis the reduced
density matrix is diagonal near a weakly avoided crossing,
and whether this is close to the diabatic $1,2$ 
basis above, or close to the $a,b$ basis of the Floquet states.
The relevant sector of the density matrix in the $1,2$ basis is given by
\begin{equation}
{\bf\rho} = \left(\begin{array}{cc}
\rho_{11} & A(\rho_{11}-\rho_{22}) \\
A^*(\rho_{11}-\rho_{22}) & \rho_{22}
\end{array}\right),
\end{equation}
with
\begin{equation}
A = \frac{1}{2d - i\Gamma /\Delta } .
\end{equation}
It is diagonal in the basis of the two states
\begin{eqnarray}
\label{diagonalbasis}
|u_{d1}\rangle & = & \left[ \frac{A}{|A|} \sqrt{1+\frac{1}{z}} \quad |u_1\rangle 
            + \sqrt{1-\frac{1}{z}} \quad |u_2\rangle \right]/\sqrt{2} \\
|u_{d2}\rangle & = & \left[-\frac{A}{|A|} \sqrt{1-\frac{1}{z}} \quad |u_1\rangle 
            + \sqrt{1+\frac{1}{z}} \quad |u_2\rangle \right]/\sqrt{2} \nonumber 
,
\end{eqnarray}
with
\begin{equation}
z = \sqrt{1+4|A|^2}
,
\end{equation}
leading to diagonal elements
\begin{eqnarray}
\rho_{d1,d1} &=& [\rho_{11} (1+z) + \rho_{22} (1-z)]/2 \\
\rho_{d2,d2} &=& [\rho_{11} (1-z) + \rho_{22} (1+z)]/2
.
\end{eqnarray}
One can check that this gives the expected result in two limiting situations:
\begin{itemize}
\item
Far from the avoided crossing, $d \gg 1$, or for  
$\Gamma  \gg \Delta$ , the effective rate $R^{\text{ac}}$ between
states $1$ and $2$ becomes small, we have $A \rightarrow 0$
and $z \rightarrow 1$, and we regain the 1,2 basis as that in which
the reduced density matrix is diagonal.
\item
For weak coupling, $\Gamma  \ll \Delta$ and near the avoided crossing, $d\ll 1$,
we have $z \gg 1$.
It is immediately clear that at 
the center, where $d \rightarrow 0$, the appropriate basis states for a
diagonal density matrix become nearly the equal admixtures of states $1$ and $2$,
which are the Floquet states $a$ and $b$. In fact, this is the case for 
all $d$ in this limit of weak coupling to the bath; for infinitesimal $\Gamma$ these
states which diagonalize $\rho$ are {\it exactly} the Floquet states.
\end{itemize}

We can readily understand these two limits in terms of the balance between energy
pumping of the system by the periodic field and relaxation by the bath.
For $\Gamma  \ll \Delta$ the near resonant periodic driving dominates, and
the Floquet states in the absence of the bath are approximate solutions, only
weakly relaxed.  In the opposite limit, $\Gamma  \gg \Delta$, those Floquet levels
are effectively broadened more by the bath interaction than their separation,
and the mixing of the diabatic states is a perturbation on their relaxation by
the bath.


\begin{thebibliography}{}
\bibitem{born} This is conventionally denoted the ``Born Approximaton" in the literature, though
it is not quite the same as the original approximation of that name in scattering theory .
\end{thebibliography}
\end{document}